\documentclass[11pt, a4paper]{article}
\usepackage{amsfonts}
\usepackage{amssymb}
\usepackage{amsmath}
\usepackage{amsthm}
\usepackage{graphicx}
\usepackage{bm}
\usepackage{url}
\usepackage{amsopn}
\usepackage{mathtools}
\usepackage{hyperref}
\usepackage{doi}
\usepackage{overpic}
\usepackage[symbol]{footmisc}
\usepackage{booktabs}
\usepackage{siunitx}
\usepackage{wasysym}
\usepackage{natbib}
\usepackage{geometry}
\usepackage{listings}
\usepackage{xcolor}
\definecolor{codegreen}{rgb}{0,0.6,0}
\definecolor{codegray}{rgb}{0.5,0.5,0.5}
\definecolor{codepurple}{rgb}{0.58,0,0.82}
\definecolor{backcolour}{rgb}{0.95,0.95,0.92}
\lstdefinestyle{mystyle}{
    backgroundcolor=\color{backcolour},   
    commentstyle=\color{codegreen},
    keywordstyle=\color{magenta},
    numberstyle=\tiny\color{codegray},
    stringstyle=\color{codepurple},
    basicstyle=\ttfamily\normalsize,
    breakatwhitespace=false,         
    breaklines=true,                 
    captionpos=b,                    
    keepspaces=true,                 
    numbers=left,                    
    numbersep=3pt,                  
    showspaces=false,                
    showstringspaces=false,
    showtabs=false,                  
    tabsize=4,
}
\begin{document}
\begin{center}
    \Large \bf Multiscale decomposition reveals predictable interannual variability and climate trends in Antarctic sea ice loss
\end{center}
\begin{center}
\begin{minipage}{0.8\textwidth}
    \begin{center}
    Peter Yatsyshin$^{*,1}$, Karl Lapo$^{*,2}$, Jonathan Smith$^{3}$ Oliver Strickson$^{1}$, Louisa van Zeeland$^{1}$, J. Scott Hosking$^{x, 1, 3}$,  J. Nathan Kutz$^{4}$
    \end{center}
\end{minipage}
\end{center}
\begin{center}
    \scriptsize{
    $*$ These authors contributed equally to this work
    ${}^1$ The Alan Turing Institute, London, UK  \\ 
    ${}^2$ Department of Cyrospheric and Atmospheric Sciences, University of Innsbruck, Innsbruck, Austria \\
    ${}^3$ British Antarctic Survey, Cambridge, UK\\
    ${}^4$ Autodesk Research, London, UK\\
    ${}^x$ Corresponding author: shosking@turing.ac.uk}
\end{center}

\begin{abstract}
Antarctic sea ice, a key geophysical system and global tipping point, has undergone unprecedented changes raising questions about how it is responding to climate change. Decades of slow expansion were replaced by a precipitous decline in 2014–2017, a subsequent apparent recovery, and a renewed collapse since 2022. We diagnosed sea ice concentration (SIC) from satellite observations with a hierarchical decomposition method based on Dynamic Mode Decomposition (DMD) that finds coherent spatiotemporal modes. We find that the 2014-2017 decline and apparent recovery are the result of interacting interannual modes. A climate change signal emerges in 2012, which becomes unambiguous by 2022 when it dominates over interannual variability. These rapid changes underscore the need for seasonal-to-annual forecasts of SIC. However, existing forecasts are subject to limited prediction horizons and high computational costs. Our predictive DMD model (\textit{IceDMD}) is regularised to prioritize the stationary spatiotemporal modes found by the decomposition. The predictive model can forecast SIC anomalies in 2023-2024 up to two years in advance, outperforming all existing approaches with the additional benefits of physical interpretability and extremely cheap computational cost. IceDMD demonstrates an unexplored direction for seasonal forecasting and this framework is likely generalizable to a range of multi-scale systems.
\end{abstract}

\section{Introduction}\label{sec1}

Antarctic sea ice impacts global and regional systems by modulating ocean circulation \cite{ferrariAntarcticSeaIce2014, rintoulGlobalInfluenceLocalized2018, joseyRecordlowAntarcticSea2024} and global energy budgets \citep{duspayevEarthsSeaIce2024}, shaping marine ecosystems \cite{atkinsonLongtermDeclineKrill2004, meredithRapidClimateChange2005, buchoveckyPotentialPredictabilitySpring2023}, acting as a global climate tipping points\citep{abramEmergingEvidenceAbrupt2025}, and buttressing of marine-terminating glaciers against ice-loss \citep{ochwatRecordGroundedGlacier2025}. Satellite records showed that a decades long trend of slow expansion of Antarctic sea ice was followed by rapid declines and substantial increases in variability \cite{parkinson40yRecordReveals2019, purichRecordLowAntarctic2023}. The first rapid decline occurred between approximately late 2014 and 2017 \citep{parkinson40yRecordReveals2019, purichRecordLowAntarctic2023}, followed by an apparent recovery, until the subsequent collapse of sea ice extent that initiated in 2021 \citep{turnerRecordLowAntarctic2022} and persists to the present \citep{abramEmergingEvidenceAbrupt2025}.  However, characterizing the drivers of the declines in SIC depends on accurately describing the time scales of the changes \citep{narayananCompoundDriversAntarctic2026}. Recent work has highlighted both the increasing role of internal variability and a likely role of climate change \citep{suryawanshiRecentDeclineAntarctic2023, bonanSourcesLowfrequencyVariability2024, espinosaUnderstandingDriversPredictability2024, narayananCompoundDriversAntarctic2026}, raising questions of whether Antarctic sea ice has entered a new climate regime \citep{purichRecordLowAntarctic2023, liuLowestAntarcticSea2023, abramEmergingEvidenceAbrupt2025}.

\begin{figure}[t]
    \centering
    \includegraphics[width=1\linewidth]{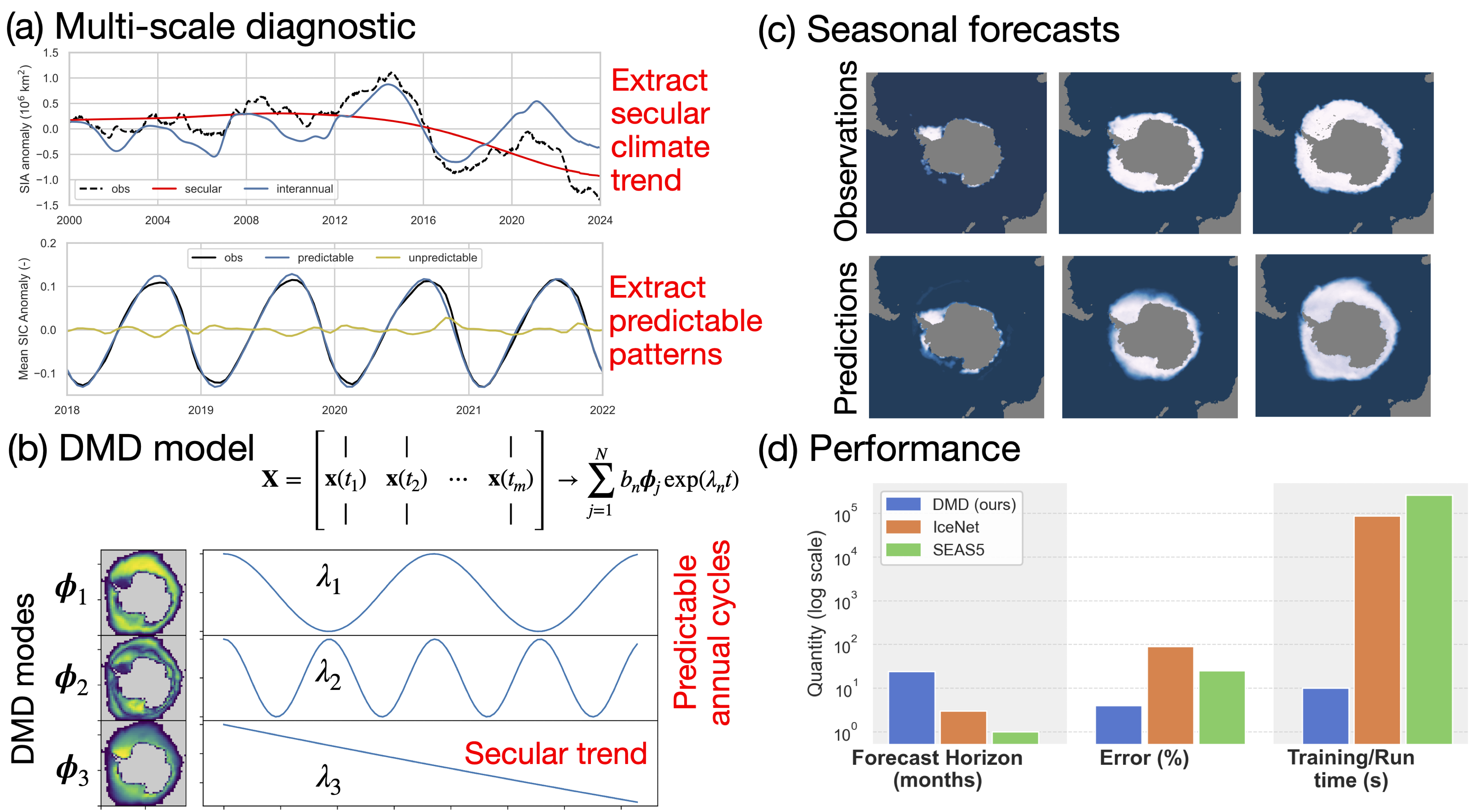}
    \caption{Our DMD framework applied to the analysis of sea ice data. (a) A multi-scale DMD decomposition of the data is used do find the predictable components of the system: repeating annual harmonics and a secular climate driven trend. (b) These patterns are used to reguralise a predictive DMD model, which models the annual harmonics and secular trend resulting in (c) accurate seasonal forecasts of sea ice. (d) The performance metrics that this forecasting framework performs better than current seasonal forecast methods at a fraction of the computational cost (see Table \ref{tab:full-comparison} for details) .}
    \vspace*{-.2in}
    \label{fig:overview}
\end{figure}

Predicting and understanding geophysical systems such as sea ice on seasonal-to-interannual (S2I) time scales (a month to a year) is essential for climate science and policy, especially as these systems potentially enter new climate regimes. This need is especially acute considering the low confidence in existing future projections of Antarctic sea ice generally owing to limited agreement between model simulations and observations, limited reliable observations on a process level, and a lack of process understanding of the substantial spread in climate simulations \citep{roachAntarcticSeaIce2020, shuAssessmentSeaIce2020}. To this end, we develop a data-driven, unsupervised, hierarchical decomposition algorithm that is capable of generating an interpretable predictive model of sea ice dynamics by disambiguating between the coherent spatiotemporal modes that are stationary and non-stationary (Fig. \ref{fig:overview}a,b), producing forecasts that outperform current state-of-the-art deep learning methods in predictive accuracy, computational efficiency and data requirements (Fig. \ref{fig:overview}c,d; Table \ref{tab:full-comparison}).

Current S2I methods for sea ice include simple statistical approaches like climatology \cite{massonnetSIPNSouthSix2023}, physically-based numerical models \cite{johnsonSEAS5NewECMWF2019, bushukSeasonalPredictionPredictability2021}, deep learning \cite{anderssonSeasonalArcticSea2021, destineAIExplainer2025}, and hybrid deep-learning/physically-based modelling \citep{gregoryAdvancingGlobalSea2026}. See \citep{massonnetSIPNSouthSix2023} for an overview of methods currently used to predict Antarctic sea ice. Deep learning, physically-based, and hybrid approaches all promise S2I forecasting on horizons of 1 to 6 months. However, these models require substantial computational resources: deep learning require large training datasets and long training times while physically-based models simulate coupled atmosphere and ocean dynamics as well as employ large ensembles (Table \ref{tab:full-comparison}). Deep-learning models also function as black boxes which limits their adoption in operational settings. Prior to the collapse of Antarctic sea ice, previous studies found that Antarctic sea ice was more predictable at S2I scales ($\approx$ 11 months) than Arctic sea ice ($\approx$1-4 months) \citep{bushukSeasonalPredictionPredictability2021}. However, subsequent studies, focusing on the period since 2017 when Antarctic sea ice may have entered a new climate regime, have found degraded S2I predictability \citep{massonnetSIPNSouthSix2023} with climate change being a possible cause \citep{liberaOceanSeaIceProcesses2022}. Thus, one of the main goals for this study is to formulate a S2I forecast for Antarctic sea ice that is applicable to emerging, novel climate states.

In this work we build a state-of-the-art S2I forecast of sea ice, called IceDMD, by diagnosing the predictable components of Antarctic sea ice.  Previous work has focused on the predictable aspects of the dynamics of the system \citep[e.g.,][]{bushukSeasonalPredictionPredictability2021, liberaOceanSeaIceProcesses2022, espinosaUnderstandingDriversPredictability2024}. In contrast, we seek the predictable patterns that can be learned using data-driven modelling. In a data-driven modelling framework, the most predictable patterns are those that repeat regularly in the historical record, i.e. they are stationary, and simple linear trends, i.e. secular trends from climate change. We use a multi-scale diagnostic \citep{lapoMethodUnsupervisedLearning2025} to disambiguate a secular climate trend signal from interannual modes of variability as well as discover the repeating, stationary patterns of predictability resulting from the annual cycle (Fig. \ref{fig:overview}a). Using this diagnostic, we strongly regularise a predictive {\em Dynamic Mode Decomposition} (DMD) model that is capable of representing these highly-predictable components of the sea ice system (Fig. \ref{fig:overview}b). This design choice results in a parsimonious predictive model that forecasts sea ice up to 2 years into the future (Fig. \ref{fig:overview}c) by avoiding the overfitting of non-stationary components of the signal. This framework yields a computationally light, transparent predictive model that forecasts Antarctic SIC more accurately and for longer prediction horizons than deep learning and physically-based coupled models, all at a fraction of the cost and with full physical interpretability (Fig. \ref{fig:overview}d; Table \ref{tab:full-comparison}).

To achieve this framework, we therefore address three primary goals: (i) understanding the drivers of the recent changes in SIC, (ii) constructing a predictive model of SIC, and (iii) providing a physically interpretable model of the SIC behavior.
This is done using the DMD algorithm which seeks to decompose time-varying data sets into a low-rank set of coherent spatiotemporal structures~\cite{kutzDynamicModeDecomposition2016}. Specifically, the temporal evolution is assumed to be approximated by a linear dynamical system $\dot{\mathbf{x}}(t) = \mathbf{A} \mathbf{x}(t)$ whose evolution dynamics are exponentials of the form

\begin{equation}
    \mathbf{\tilde{x}}(t) = \sum_{j=1}^N b_j \boldsymbol{\phi}_j \exp(\lambda_j t)
    \label{eq:DMDsolution}
\end{equation}

with the imaginary component of the exponent $\lambda_j = \mu_j + i\omega_j$ modelling periodic temporal oscillations ($\omega_j$) and the real component ($\mu_j$) modelling exponentially growing or decaying phenomena.  The $\mathbf{\tilde{x}}$ refers to an approximation of the original data or state-space, $\mathbf{x}$. The $\boldsymbol{\phi}_j$ are the coherent spatial modes. For the data considered here, these will be Antarctic spatially coherent (modal) structures dominating sea ice activity. The parameter $b_j$ is the weight of each DMD mode when linearly superimposing for reconstructing the overall sea-ice concentration. Finally, $j$ specifies the rank of a mode, up to a total rank of $N$. A DMD mode consists of conjugate pairs of both $\phi_j$ and $\omega_j$. When selecting an odd rank a DC component emerges which describes non-oscillatory behavior. Thus, in our application of DMD we discover dominant modal features that have specific temporal dynamics that model sea ice dynamics.

In the first part of this work we diagnose the drivers of the recent changes SIC using a multi-scale DMD variant called {\em multiresolution coherent spatiotemporal scale separation} (mrCOSTS) \citep{lapoMethodUnsupervisedLearning2025}. The mrCOSTS algorithm is an unsupervised, hierarchical decomposition based on DMD that finds coherent spatial patterns described by narrow temporal frequency bands. mrCOSTS robustly diagnoses complex multi-scale and non-stationary signals that challenge other decomposition methods and has been employed to understand a range of complex geophysical systems \citep{lapoMethodUnsupervisedLearning2025, uchidaDynamicModeDecomposition2025, lapoScaleAwareEvaluationComplex2025}.  In the second part, we construct a seasonal forecasting framework in which the multi-scale diagnostic informs a strongly regularised predictive DMD model. The mrCOSTS diagnostic is used to discover the predictable modes of the SIC signal by disambiguating between the coherent spatiotemporal modes that are stationary and non-stationary, and therefore between modes that are more and less predictable, respectively. We regularise the predictive DMD model with a low-rank approximation that corresponds to the large-amplitude, stationary patterns found by the multi-scale decomposition. The predictive model is then evaluated and compared against other methods. Finally, the overall framework, a multi-scale diagnostic used to inform a predictive model, is interpreted within the context of forecasting and diagnosing Antarctic sea ice and we present the framework as a potentially generalizable method.

\begin{table}[h]
\centering
\footnotesize
\caption{Comparison of methods for S2I forecasting of sea ice. $^{*}$ indicates short-term forecasting tools (forecast horizon less than 1 month), which are only included for comparison of model cost. For physically-based systems, training data refers to the approximate size of the reanalysis related to sea ice and training/run time refers to the amount of time required to generate a forecast. Size of training data estimated based on methods descriptions and is only approximate. $^x$ indicates an Arctic sea ice forecasting systems. Hybrid models only describe the machine learning costs but also include the cost of running a fully coupled model. $^t$ indicates a machine learning emulator.}
\label{tab:full-comparison}
\begin{tabular}{p{3.cm}p{1.cm}p{1.5cm}p{1.3cm}p{1.5cm}p{2cm}p{1.5cm}p{1.25cm}}
\toprule
Method
  & Model Type
  & Forecast Horizon
  & Training Data
  & Training/ Run Time
  & Model Size
  & Error Metric
  & Error Value \\
\midrule
IceDMD
  & Data-driven
  & 2 years
  & $\mathcal{O}$(500~MB)
  & Seconds (1 CPU)
  & 5 modes
  & MAE
  & $\sim$4\% \\[4pt]
Hybrid SPEAR \citep{gregoryAdvancingGlobalSea2026}
  & Hybrid
  & 12 months
  & $\mathcal{O}$(10~GB) (2619 samples)
  & $\mathcal{<1~h}$ on 1 GPU
  & $\mathcal{O}(10^5)$ for deep learning
  & RMSE
  & 10-25\% \\[4pt]
IceNet$^{x}$\citep{anderssonSeasonalArcticSea2021}
  & Deep-learning
  & 1--6 months
  & $\mathcal{O}$(1-10~TB)
  & $\sim$1 d/member (GPU) $\times$ 25
  & $\sim$1.1$\times10^{9}$ params
  & Integrated Ice Edge Error (IIEE)
  & 93--96\% \\[4pt]
SEAS5$^{x}$\citep{johnsonSEAS5NewECMWF2019}
  & Physics-based
  & 7-13 months
  & $\mathcal{O}$(1~TB)
  & $\mathcal{O}$(1-5 days)
  & 51 member ensemble, fully coupled, 0.25$^{\circ}$ ocean, 36~km$^2$ atmosphere
  & RMSE
  & 10-25\% \\[4pt]
DestinE ML Sea Ice$^{*,x}$ \citep{destineAIExplainer2025}
  & Deep-learning
  & 10 days
  & $\mathcal{O}$(100~GB)
  & $\mathcal{O}$(2 days) (16 GPUs)
  & $\mathcal{O}$(10$^7$-10$^8$) params
  & IIEE
  & 15\%-20\% \\
FloeNet Sea Ice Emulator$^{t}$ \citep{gregoryFloeNetMassconservingGlobal2026a}
  & Deep-learning
  & N/A (climate emulator)
  & $\mathcal{O}$(100~GB)
  & $\mathcal{O}$(5 days) (4 GPUs)
  & $\mathcal{O}$(10$^6$) params
  & IIEE
  & 15\%-20\% \\

\bottomrule
\end{tabular}
\end{table}

\section{Results}\label{sect:results}

\subsection{Multi-scale diagnosis of recent changes}
\begin{figure}
    \centering
    \includegraphics[width=1\linewidth]{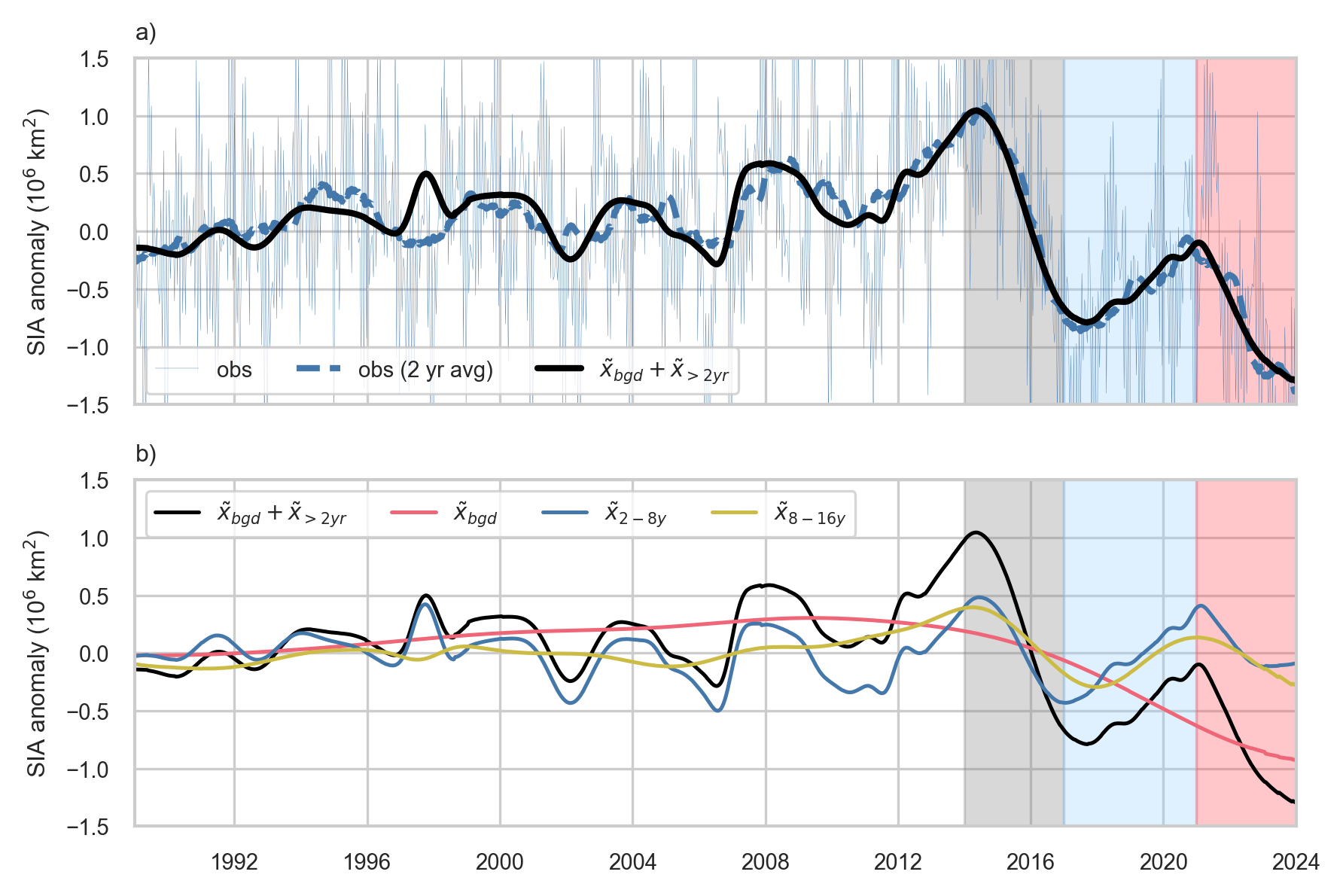}
    \caption{(a) The observed monthly Sea Ice Area (SIA) anomaly and its 24-month rolling average is compared to the sum all interannual modes and the slow background mode. (b) The individual interannual modes resolved by mrCOSTS that contribute to the black line in (a) are shown. Intervals of are highlighted: 2014-2017 decline in grey, 2017-2021 recovery in blue, and the 2021-present decline in red. }
    \label{fig:interannual-variability}
\end{figure}

mrCOSTS was used to diagnose coherent spatiotemporal modes in the satellite observed SIC \citep[see Methods;][]{lavergneVersion2EUMETSAT2019}. mrCOSTS modes with a time scale longer than two years were aggregated into three bands with time scales of:\\

\noindent
(1) two to eight years corresponding to an ENSO-like band \citep[$\tilde{\mathbf{x}}_{2-8}$,][]{wangImpactsCombinedSAM2023, bonanSourcesLowfrequencyVariability2024}, 

\noindent
(2) eight to sixteen years corresponding to a Decadal Oscillation \citep[$\tilde{\mathbf{x}}_{8-16}$,][]{meehlAntarcticSeaiceExpansion2016, liuDecadalOscillationProvides2023, bonanSourcesLowfrequencyVariability2024},

\noindent 
(3) a ``background" mode, $\tilde{\mathbf{x}}_{bgd}$, that contains all time scales longer than 20 years.\\

\noindent
As noted previously, $\tilde{\mathbf{x}}(t)$ indicates a modal approximation of the given state-space ${\mathbf{x}}(t)$ with $\tilde{\mathbf{x}}_n$ indicating the $n$th mode. We interpret the background mode as being a combination of climate change and multi-decadal time scales since mrCOSTS cannot distinguish between modes with time scales close to the duration of the data or longer and secular trends, i.e. the long-term consistent upward or downward movement in data over extended periods. In mrCOSTS secular trends emerge as the lowest frequency components in $\tilde{\mathbf{x}}_{bgd}$. The sum of the these terms determines our state space approximation

\begin{equation}
   \tilde{\mathbf{x}}(t) = \tilde{\mathbf{x}}_{bgd} + \tilde{\mathbf{x}}_{2-8} + \tilde{\mathbf{x}}_{8-16}
\end{equation}

closely follows the rolling 2-year average of the observed SIA anomalies (Fig. \ref{fig:interannual-variability}a; see Methods), highlighting how the mrCOSTS decomposition cleanly diagnoses the observed data at interannual time scales.

These decomposed bands allow us to robustly diagnose the recent recent sea ice anomalies. The interannual modes, $\tilde{\mathbf{x}}_{2-8}$ and $\tilde{\mathbf{x}}_{8-16}$ were small amplitude at the beginning of the record and gradually ``turned on" after 2000 with both modes reaching their largest amplitude after 2012. The interaction of these modes both in a negative phase characterized the collapse of SIC between 2014-2017 (grey region in Fig. \ref{fig:interannual-variability}b). The following apparent recovery (2017-2021; blue region in Fig. \ref{fig:interannual-variability}a) was a result of both modes returning to a positive phase. Finally, both modes entered effectively a neutral phase by 2024 (2021-present; red region in Fig. \ref{fig:interannual-variability}b). 

The background mode, $\tilde{\mathbf{x}}_{bgd}$ shows a slow increase until approximately 2012 at which point it starts slowly decreasing, followed by a more substantial decreases after 2017. Notably, between 2014-2017 $\tilde{\mathbf{x}}_{bgd}$ does not contribute to a negative SIA anomaly. Instead, $\tilde{\mathbf{x}}_{bgd}$ contributes a significantly negative SIA anomaly after 2021, when the trend in this term exceeds the amplitude of the interannual oscillations in $\tilde{\mathbf{x}}_{2-8}$ and $\tilde{\mathbf{x}}_{8-16}$, which is $\approx$0.5 10$^6$ km$^2$. We estimate the decline from $\tilde{\mathbf{x}}_{bgd}$ to be responsible for the loss of $\approx$0.75 10$^6$ km$^2$ since 2012 or $\approx$10$^6$ km$^2$ as measured from the peak contribution of this mode, corresponding to the observed anomaly in SIA in 2024 (Fig. \ref{fig:interannual-variability}a).

\subsection{Climate change signal in the background mode}
\begin{figure}
    \centering
    \includegraphics[width=1\linewidth]{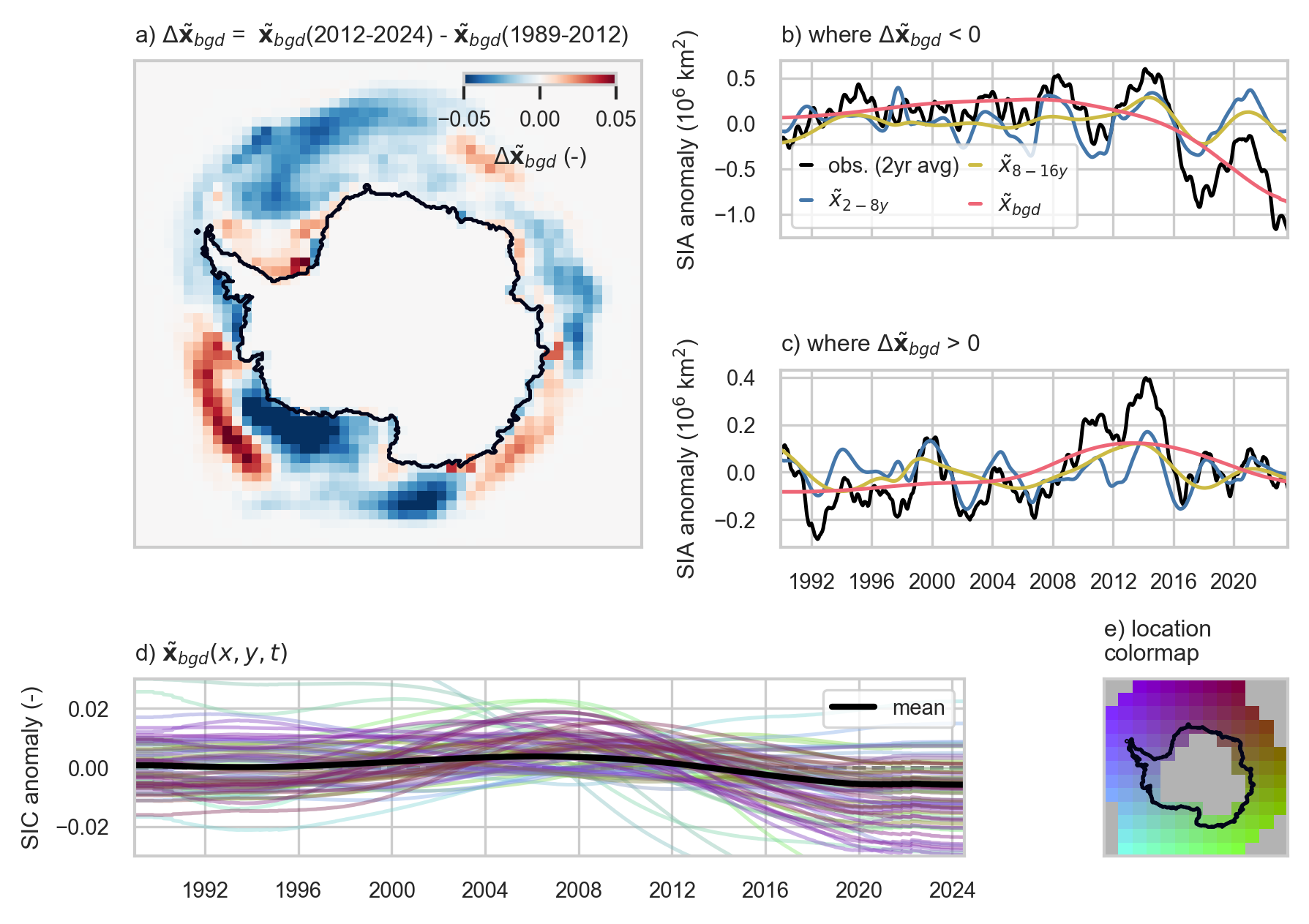}
    \caption{a) The spatial pattern of changes in $\tilde{\mathbf{x}}_{bgd}$ from 2012-2024 relative to a baseline period from 1989 and 2012. (b) SIA anomalies from each of the mrCOSTS bands at pixels where $\tilde{\mathbf{x}}_{bgd}$ decreased and (c) $\tilde{\mathbf{x}}_{bgd}$ increased relative to the baseline period are shown. A 12-month rolling average was applied to the observed SIA anomalies. (d) The anomaly in SIC from $\tilde{\mathbf{x}}_{bgd}(t)$ is shown for each pixel corresponding to the color mapped in (e). For (d) $\tilde{\mathbf{x}}_{bgd}(t)$ was aggregated to 4x4 pixels for visual clarity. Note that subplots (a) and (d) show SIC while (b) and (c) use SIA.}
    \label{fig:mrcosts-background-mode}
\end{figure}

The difficulty in interpreting $\tilde{\mathbf{x}}_{bgd}$ is the inability to distinguish between multi-decadal motions known to exist in SIC data \citep{moriokaAntarcticSeaIce2024, narayananCompoundDriversAntarctic2026} and any secular contribution from a climate change signal. Resolving $\tilde{\mathbf{x}}_{bgd}$ into its constituent parts would require a longer historical record of SIC data than is available, so we are forced to treat these unresolved low-frequency components cumulatively as the background in mrCOSTS analysis.

The secular decrease in the background mode contributed to negative SIA anomalies that exceed the magnitude of the interannual modes both regionally and locally (Fig. \ref{fig:mrcosts-background-mode}). The spatial pattern of how $\tilde{\mathbf{x}}_{bgd}$ changed relative to a baseline period prior to 2012 reveals that $\tilde{\mathbf{x}}_{bgd}$ contributed to a decrease in SIA across nearly the entire region (blue regions in Fig. \ref{fig:mrcosts-background-mode}). A small increase in SIA relative to this same baseline was found for the margins of the Amundsen Sea, near the coast in the Weddell Sea, and the D'Urville Sea (red regions in Fig. \ref{fig:mrcosts-background-mode}a). The gain in SIA in $\tilde{\mathbf{x}}_{bgd}$ was reversed by 2024, consistent with a multi-decadal oscillation superimposed on a decreasing secular climate trend. The rest of the Southern Ocean experienced declines in the $\tilde{\mathbf{x}}_{bgd}$, with the largest local losses occurring in the Amundsen and Ross Seas nearer to Antarctica and in the Somov Sea.

The decline in $\tilde{\mathbf{x}}_{bgd}$ starting in approximately 2012 does not have a historical analogue within the observed record (Fig. \ref{fig:interannual-variability}b). The spatial patterns of the decline in $\tilde{\mathbf{x}}_{bgd}$ appear to support the idea that the decrease since 2012 is the result of a novel climate signal. The time series of $\tilde{\mathbf{x}}_{bgd}$ across all pixels reveals multi-decadal scale oscillations at individual locations but a secular, spatially extensive decline that initiates in 2012 clearly dominates the regional behavior (Fig. \ref{fig:mrcosts-background-mode}d). 

\subsection{The Predictive DMD Model}

\subsubsection{Discovering Stationary Patterns}

\begin{figure}
    \centering
    \includegraphics[width=0.8\linewidth]{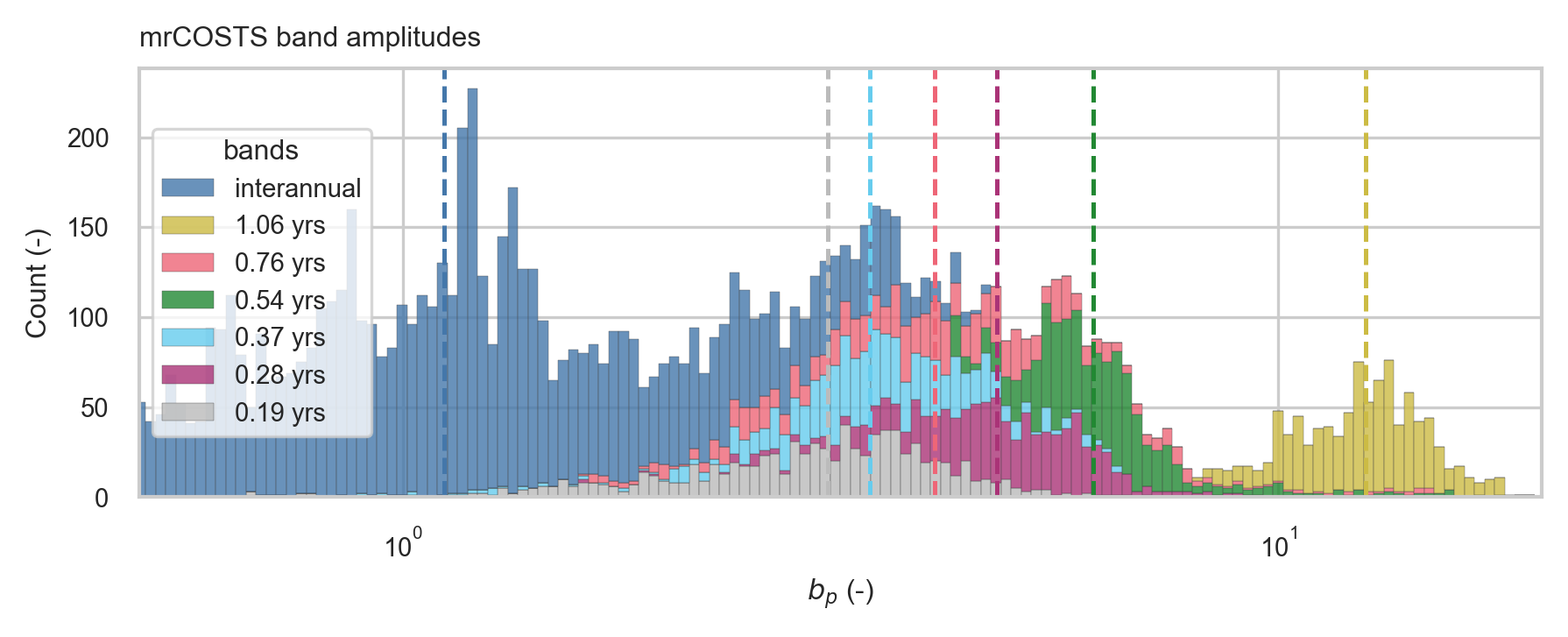}
    \caption{Histograms of each band's amplitudes, $b_p$ (Eq. \ref{eq:summed-amplitude}), with the vertical dashed lines indicating the band median. The bands with a time scale longer than a year are grouped together for clarity.}
    \label{fig:mrcosts-bands}
\end{figure}

DMD struggles to characterize non-stationary patterns, including spatial patterns that change through time or that turn on or off \citep{kutzDynamicModeDecomposition2016}. For this reason we seek to regularise our predictive DMD model to fit only patterns that it can be expected to accurately model. The multi-scale decomposition from mrCOSTS allows us to investigate which coherent spatiotemporal modes are characterized by stationary patterns in both time and space. 

\begin{figure}
    \centering
    \includegraphics[width=.8\linewidth]{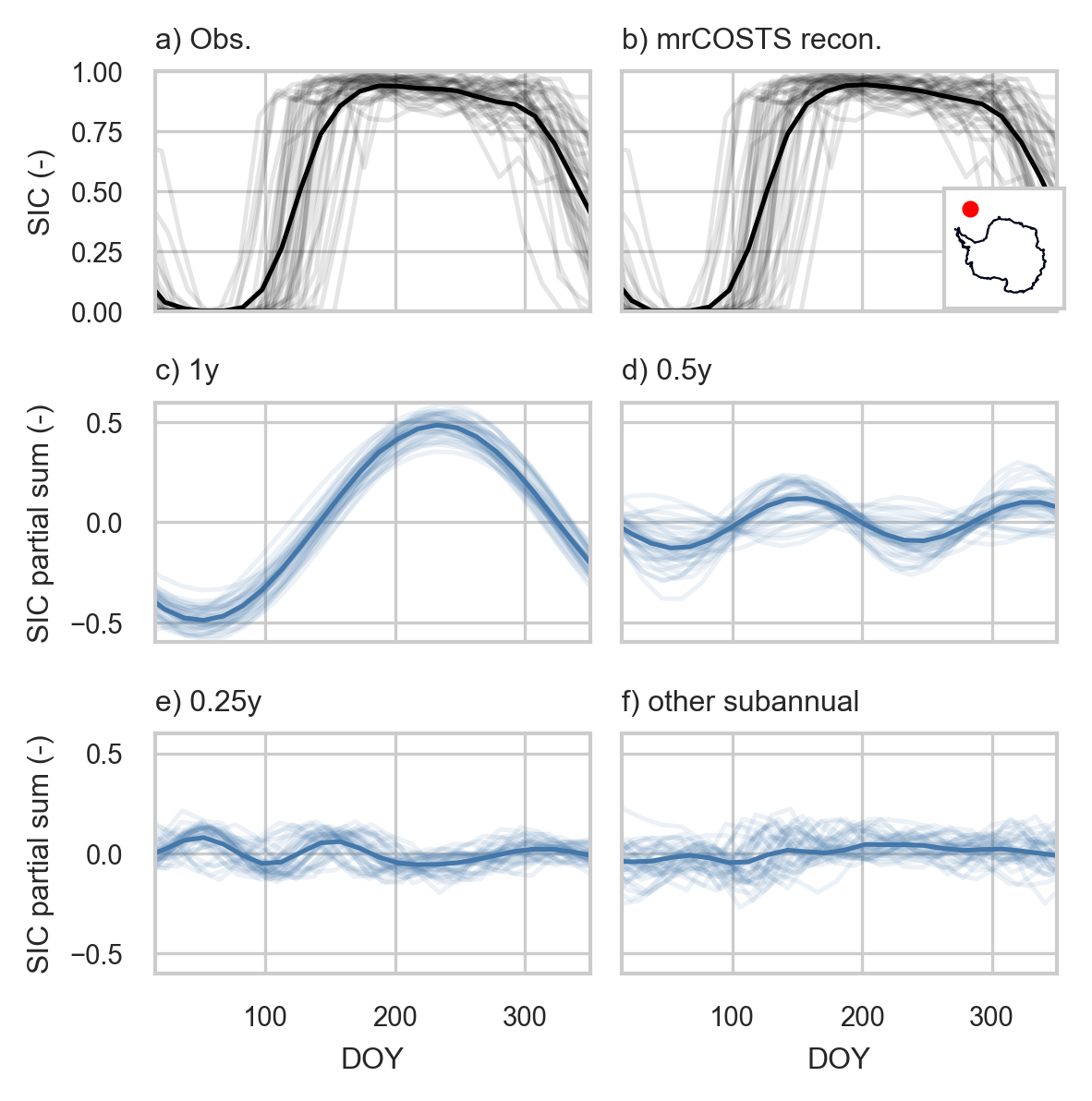}
    \caption{mrCOSTS diagnosis of SIC at a point in the Weddell Sea (see inset map) by day of year (DOY) for all years considered. (a) and (b): observed and reconstructed SIC for each year in the dataset. (c)-(f): Reconstructions of annual and subannual mrCOSTS bands, as given in equation \eqref{eq:summed-amplitude}. Faint curves represent individual years, and dark curves represent their annual composite means.}
    \label{fig:mrcosts-point-diagnosis}
\end{figure}

Unsurprisingly, the decomposition found that the annual and half-year modes dominate the temporal dynamics of SIC. The annual mrCOSTS mode had the largest amplitude in the mrCOSTS decomposition ($b_p$ in Eq. \ref{eq:summed-amplitude}) followed by the half-year mode and quarter-year modes (Fig. \ref{fig:mrcosts-bands}). The annual and half-year modes are characterized by consistent patterns year-to-year, as illustrated using the reconstruction of each band, $\tilde{\mathbf{x}}_p$, at an individual point in the Weddell Sea (Fig. \ref{fig:mrcosts-point-diagnosis}; Fig. \ref{fig:overview}a shows the area-averaged SIC for reference). The mrCOSTS reconstruction captures the annual cycle of the observed SIC (compare Fig. \ref{fig:mrcosts-point-diagnosis}a to b). This reconstruction is largely driven by the annual mode (Fig. \ref{fig:mrcosts-point-diagnosis}c), which is remarkably consistent across years. The half-year mode shows a pattern that also repeats between most years, but a phase shift relative to the average pattern emerges in some years (Fig. \ref{fig:mrcosts-point-diagnosis}d). The remaining higher-frequency modes (Fig. \ref{fig:mrcosts-point-diagnosis}e and f) are clearly non-stationary between years and even within a given year. For instance, the 0.25-year band does not regularly oscillate across the year, but turns on and off, with the largest amplitudes in the spring and its timing varying between years. Similar observations can be made about 0.37- and 0.19-year bands, which are summed in Fig. \ref{fig:mrcosts-point-diagnosis}f for brevity. Although we illustrated the non-stationarity of high-frequency modes at a single location, the pattern holds across the entire region (Fig. \ref{fig:overview}a).

While the annual and half-year bands reconstruct the majority of the annual cycle's amplitude, the highly irregular high-frequency modes characterize the rapid changes in the shoulder seasons. In particular, the sum of the annual and half-year modes fail to capture the rapid onset of freezing, which are characterized by the higher frequency modes. However, these modes contribute relatively little to the overall signal while also being simultaneously non-stationary and thus much harder to predict.

\begin{figure}
    \centering
    \includegraphics[width=1\linewidth]{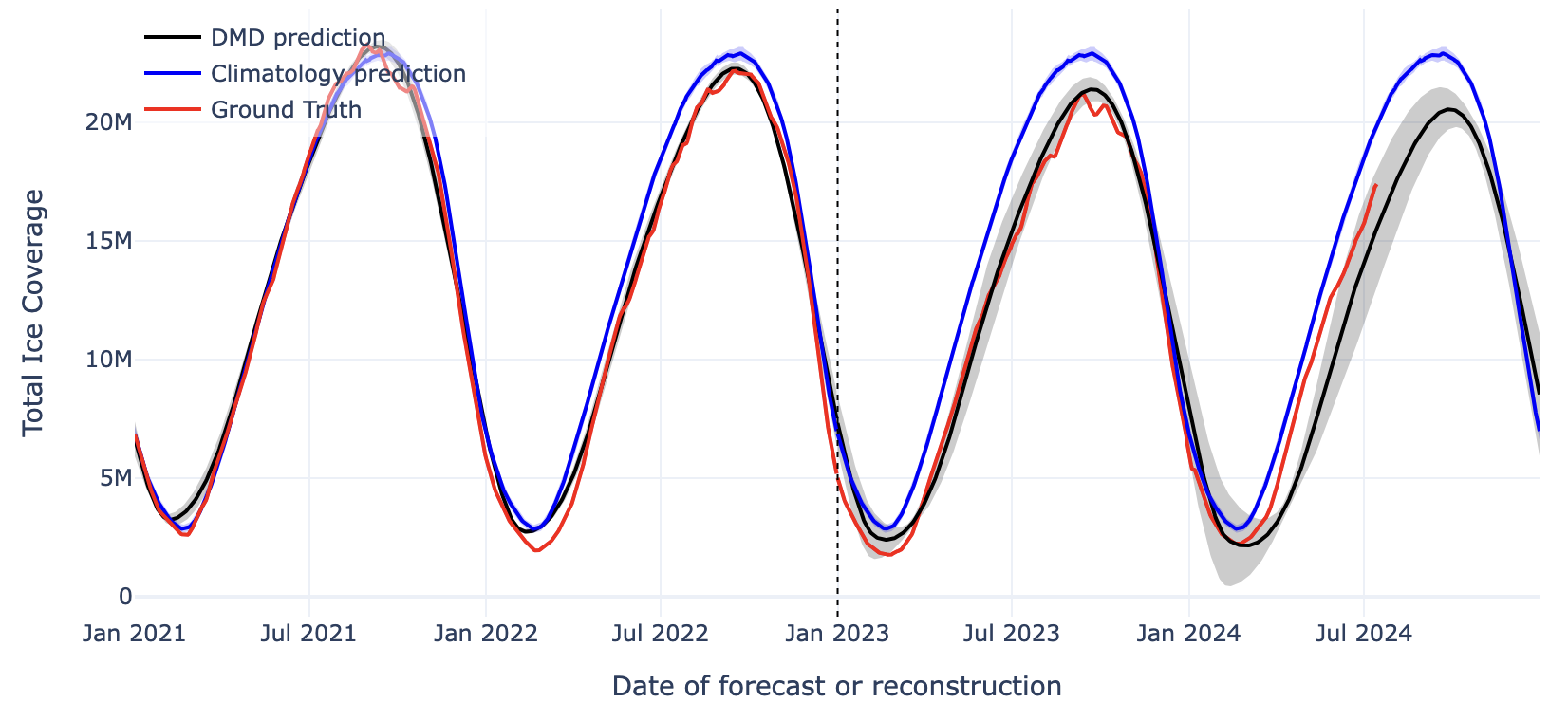}
    \caption{Weekly-averaged observed, reconstructed, and predicted SIA, with the forecast beginning on Jan 1, 2023 (vertical dashed line). Also shown are $2\sigma$ uncertainty bands obtained by the bagging operation when fitting the predictive DMD model (see Methods).}
    \label{fig:Total-ice-pred}
\end{figure}

\subsubsection{Regularising the Predictive Model}

The DMD framework fits the dominant, spatially coherent patterns that describe SIC. We regularise the predictive DMD model through rank selection, eigenvalue constraints, and a restricted training window guided by the mrCOSTS diagnosis. The high amplitude, stationary annual and half-annual modes that dominate SIC behaviour inform the rank hyperparameter choice in the predictive DMD model by limiting the model's rank to $j=5$ (Eq. \ref{eq:DMDsolution}), corresponding to four complex-conjugate modes and one DC component. Attempting to predict the higher frequency subannual modes with DMD, for instance by using a larger rank, is unjustified given their non-stationary characteristics. The emergence of a negative secular trend that dominates over interannual variability motivates using an odd rank after 2021 when the apparent secular trend dominates (Fig. \ref{fig:interannual-variability}b). This regularisation allows the predictive DMD model to estimate the influence of this apparent climate signal through the non-oscillatory temporally decaying DC mode. Thus, the predictive DMD model is regularised to have a rank of 5, 2 oscillatory modes and 1 decaying, non-oscillatory mode.

We prescribed a training window of two years prior to the forecast for the predictive model. This regularisation of a two year training window allows the DMD model to capture the relatively stationary annual and half-year modes, with at least two full cycles of each. Longer training windows contaminate the SIC signal with the lower frequency interannual modes which are highly non-stationary (Fig. \ref{fig:interannual-variability}b) and thus less predictable with the DMD model.

\subsubsection{Evaluating the predictive model}

\begin{figure}[h]
    \centering
    \includegraphics[width=0.6\linewidth]{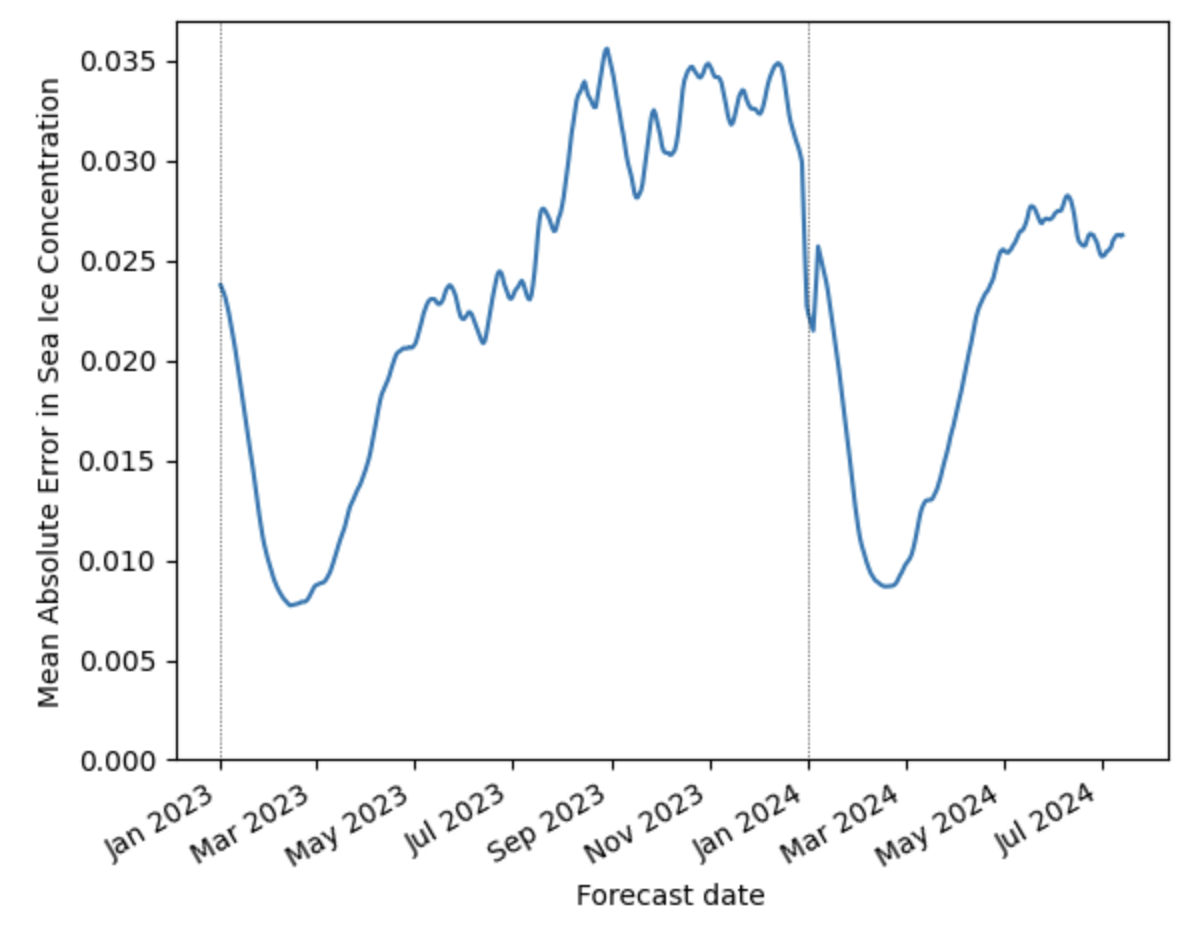}
    \caption{Mean  absolute error of predicted daily SIC configurations. Vertical lines are visual guides drawn at January 1, 2024 and January 1, 2025.}
    \label{fig:dmd-forecast-eval}
\end{figure}

We demonstrate the predictive DMD model by forecasting the 2023 and 2024 trends in SIA (Fig. \ref{fig:Total-ice-pred}) up to two years in advance. The training period spans January 1, 2021 to January 1, 2023. The DMD model predicted SIA across the domain (Fig. \ref{fig:Total-ice-pred}) with a MAE of $\approx$4\% up to 2 years into the future (Fig. \ref{fig:dmd-forecast-eval}). The predictive model captures modes whose periodicity is commensurate with the annual and half-year modes within the 2-year horizon (Fig. \ref{fig:overview}b). The DMD prediction also outperforms the SIA climatology over the prediction window specifically by modelling the secular trend in SIC known to exist during this period (Fig. \ref{fig:interannual-variability}b and \ref{fig:overview}b).

\begin{figure}
    \centering
    \includegraphics[width=0.9\linewidth]{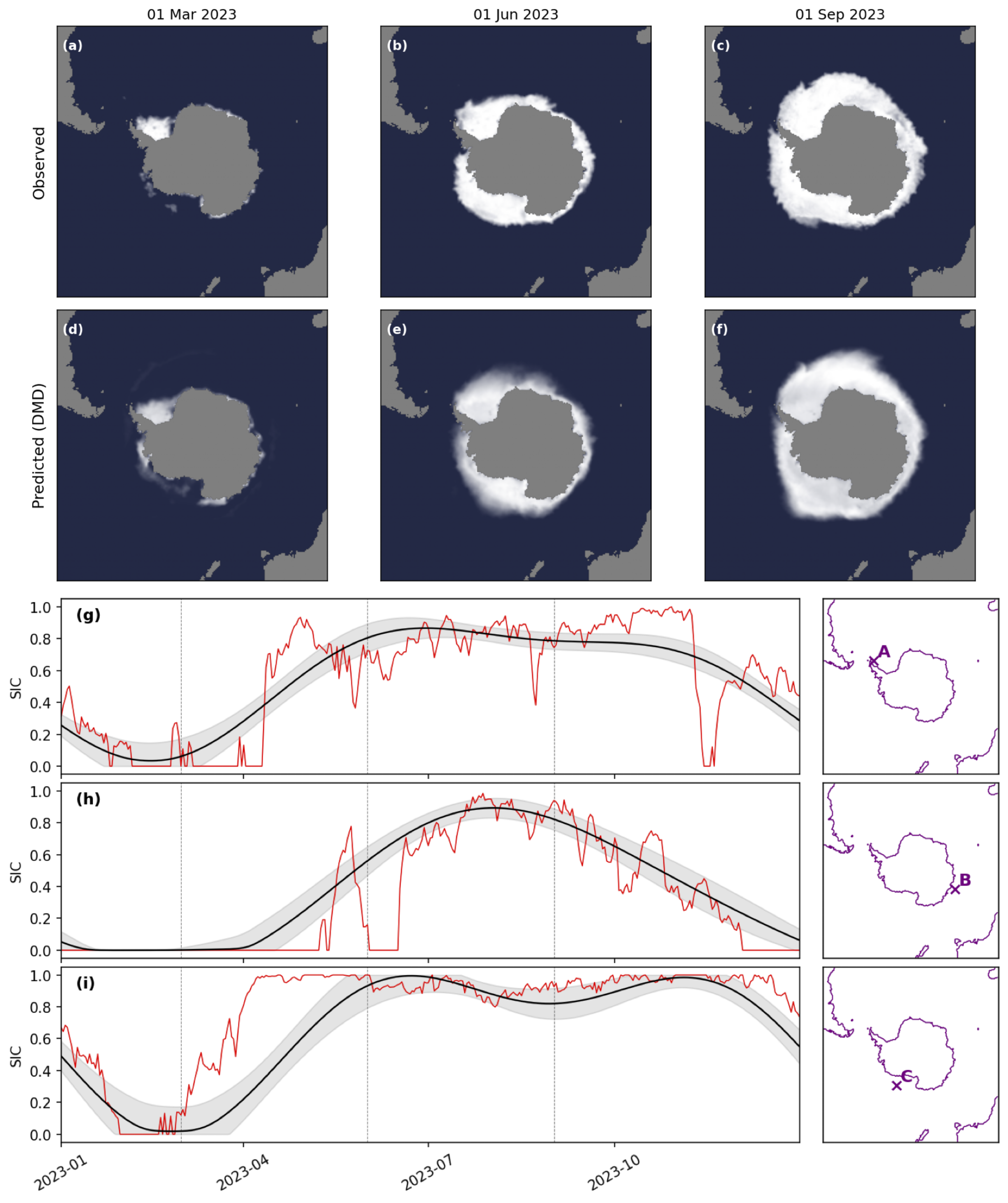}
    \caption{(a-c) Observed and (d-f) predicted spatial patterns of SIC corresponding to the minimum (a,d), freezing phase (b,e), and maximum in SIA (c,f). (g-i) Time series of observed and predicted SIC at three locations (see corresponding maps) during 2023 with observations in red, DMD prediction in black, and $2\sigma$ uncertainty bands in grey. Vertical dashed lines correspond to the times shown in (a-f).}
    \label{fig:pred-dmd-spatial+point}
\end{figure}

The predictive DMD model misses aspects of the freeze and melt seasons. The errors are highest in September, at the height of the freeze season but are the smallest in the early spring (Fig. \ref{fig:dmd-forecast-eval}). This result is consistent with the mrCOSTS decomposition, which shows that the excluded high frequency modes contribute less to the signal in the second half of the year (Fig. \ref{fig:mrcosts-point-diagnosis}e,f). The predictive DMD model also accurately predicts the spatial patterns of SIC (Fig. \ref{fig:pred-dmd-spatial+point}a-f). Generally the DMD prediction is smooth relative to the observations while still capturing the overall trends of the observations. The predicted SIC spatial patterns qualitatively agree well with the observed configurations but do not reach the saturated SIC values that were observed near the SIC maxima (Fig. \ref{fig:pred-dmd-spatial+point}c,f). Similarly, the predicted time series clearly miss the higher-frequency and finer-scale components of SIC at any given location while still capturing the overall trend \ref{fig:pred-dmd-spatial+point}g-i). The uncertainty band, representing 2$\sigma$ uncertainty from the bagging operation in the DMD fitting, encompasses the observed SIA across the entire region (Fig. \ref{fig:Total-ice-pred}). However, the DMD ensemble is under dispersive, with the ensemble spread being poorly correlated with the model error, as well as underestimating uncertainty at any given location, especially in the shoulder seasons (Fig. \ref{fig:pred-dmd-spatial+point}).

\subsubsection{Comparison to Other Seasonal Forecasts}\label{sect:comparison}

The predictive DMD model is dramatically cheaper in terms of computation expense and training data needs than other methods for the S2I forecasting of sea ice (Table \ref{tab:full-comparison}). Deep-learning approaches have generally substantial training data needs, e.g. IceNet pre-trained on 2220 years of CMIP6 data \citep{anderssonSeasonalArcticSea2021} and the hybrid SPEAR model required training on a 30 member ensemble of coupled ocean-atmosphere model output for a 35 year interval \citep[][Table \ref{tab:full-comparison}]{gregoryAdvancingGlobalSea2026}. Similarly, physically-based, deep-learning, and hybrid models are comparatively computationally expensive to either train (deep-learning and hybrid approaches) or run (physically-based models). Finally, these models are generally large, e.g., IceNet deploys 25 neural networks with a combined 1.1 billion parameters \cite{anderssonSeasonalArcticSea2021}. Our approach, by contrast, uses just five DMD modes with clear physical interpretations \citep{kutzDynamicModeDecomposition2016}, and trains in seconds on a single CPU using only two years of satellite SIC observations (Table \ref{tab:full-comparison}). Overall, the DMD model uses approximately 10$^3$ times less training data, trains 10$^4$ times faster, and contains $\approx$10$^8$ fewer parameters (Table \ref{tab:full-comparison}; Fig. \ref{fig:overview}d). 

Despite the lightweight nature of the DMD model it also has strong predictive skill over longer time horizons than other seasonal forecasts (Table \ref{tab:full-comparison}; Fig. \ref{fig:overview}d). Specifically, the DMD model achieves prediction horizon that far exceed those achieved by deep learning, comparing for instance 4-6 months for the hybrid SPEAR approach and $\approx$2 years for the predictive DMD. However, a quantitative comparison is difficult to make as a consequence of varying error metrics across studies.

The predictive model also outperforms climatology, which is generally the baseline for longer time horizon S2I forecasts \citep{massonnetSIPNSouthSix2023}. Although the climatology model captures the strong seasonality of the data, it obviously cannot account for changes in climate trends and interannual variability. It is for this reason the DMD model is able to outperform climatology specifically in the period for which Antarctic sea ice is entering a new climate driven state. Finally, it has been shown that statistical models generally out-perform physically-based and deep-learning models for forecasting Antarctic sea ice, but with the specific drawback of not providing uncertainty estimates \citep{massonnetSIPNSouthSix2023}. The predictive DMD model naturally provides these uncertainty estimates through the bagging operation (see Methods), however, these uncertainty estimates are underestimated.

\section{Discussion}\label{discussion}

This study consisted of two parts. In the first, we performed a multi-scale decomposition with the mrCOSTS method to understand the spatiotemporal features of recent changes in Antarctic sea ice. In the second part, the multi-scale decomposition was used to strongly regularise a predictive DMD model, enabling it to achieve season-to-annual predictions that far exceed the current state of the art.

The distinction between interannual variability and secular climate trends is critical to understanding recent Antarctic sea ice behaviour. Recent work has highlighted that mechanisms interacting on multiple interacting time scales can explain the recent collapse of Antarctic sea ice \citep{narayananCompoundDriversAntarctic2026}. This interpretation is consistent with our mrCOSTS analysis of the 2014-2017 collapse but not with the low SIC anomalies after 2021 which were not associated with interannual modes of variability. We instead argue that Antarctic sea ice has not entered an abrupt new regime, but rather that the secular trend from climate change has become large enough to dominate the SIC anomalies over other sources of variability, which has been unclear previously \citep{abramEmergingEvidenceAbrupt2025, narayananCompoundDriversAntarctic2026}. Further investigation is necessary to properly quantify the emerging secular trend, especially in light of the low sea ice extent observed since 2024. Our interpretation of the mrCOSTS analysis attributes these record low SIC primarily to the continued decline of the background mode with interannual variability playing a secondary role, somewhat at odds with recent mechanistic explanations \citep{narayananCompoundDriversAntarctic2026}. Future work will focus on incorporating more recent data and further refining the physical role of the interannual modes and secular trend.

Current diagnostic methods impose constraints that obfuscate the complex multi-scale dynamics of these kinds of geophysical systems. For instance, Principal Component Analysis and related modal methods assume fixed spatial patterns, break correlations between time and space, and cannot effectively represent non-stationary events \citep{kutzDynamicModeDecomposition2016, tairaModalAnalysisFluid2017}. Time-frequency approaches, such as wavelets and spectral methods, can diagnose the time scales present, but they can only operate along a single dimension (e.g., time but not space). Finally, normal DMD without windowing cannot effectively represent non-stationary processes \citep{kutzDynamicModeDecomposition2016, MultiresolutionDynamicMode}. Antarctic sea ice is explicitly multi-scale and non-stationary (Fig. \ref{fig:interannual-variability}), which creates a fundamental challenge when employing these classical approaches. mrCOSTS was explicitly designed to find coherent spatiotemporal modes in non-stationary signals, which is what enabled in this analysis to disambiguate between interannual variability and a secular climate trend. We highlight here the general need to adopt methods designed to diagnose these kinds of multi-scale dynamics, of which mrCOSTS is one.

The multi-scale diagnosis of SIC motivates what is perhaps the most consequential design choice in our framework: the extremely low rank of the predictive DMD model. Creating a predictive DMD model required stringent regularization such that the model fits the predictable modes of the sea ice system. The key hyperparameter to regularise is the rank to employ in the predictive DMD model (see Methods). The DMD rank represent a trade-off between capturing the predictable coherent spatiotemporal modes and accurately reconstructing the non-stationary components of the training data. A higher-rank DMD model would better fit the non-stationary modes in the training data but would be unable to generalise them for future conditions because those modes lack temporal coherence. DMD models can reconstruct features such as transience with interfering modes but these modes are not physical nor can they be projected into the future \citep{MultiresolutionDynamicMode, ferreNonStationaryDynamicMode2023, lapoMethodUnsupervisedLearning2025}. 

We highlight here the critical importance of parsimonious rank selection in order to model physically meaningful and predictable modes, as this point does not appear to have been made previously. The parts of the system driven by consistent and regular outside forcing, in this case the regular patterns of orbital forcing of solar irradiances and secular trends from climate change, will tend to generate the most predictable patterns. The targeted regularization of the predictive model to discover these predictable parts of the system appears to be an under exploited strategy for building predictive models generally, e.g. for S2I forecasting, and may explain the overall improved performance of statistical models relative to more expensive frameworks \citep{massonnetSIPNSouthSix2023}. We do not believe this point has been well-acknowledged previously for its practical implications in data-driven modelling or machine learning frameworks.

We speculate that the non-stationary high-frequency modes may be a complimentary explanation to the apparent degraded S2I forecasts in the Antarctic winter. Previous work has focused on specific dynamical processes that are harder to model in the Antarctic summer \cite[e.g., mixed layer shoaling, ice-albedo feedback, vertical mixing of the ocean, entrainment of warm water that has not been influenced by sea ice][]{liberaOceanSeaIceProcesses2022, massonnetSIPNSouthSix2023}. As a compliment to these analyses we additionally diagnose that the period between approximate January and April is characterized by a stronger contribution from the irregular, non-stationary high-frequency modes, which may contribute to the difficulty of predicting through this transition. Second, we demonstrated the emergence of a secular trend and interannual variability, especially since 2012. Prior to this period Antarctic sea ice was remarkably stable in its spatiotemporal patterns (Fig. \ref{fig:interannual-variability}). This stability of the system may explain the prior discovery of highly predictive SIC \citep{bushukSeasonalPredictionPredictability2021} that has not been present in more recent forecasts \citep{liberaOceanSeaIceProcesses2022, massonnetSIPNSouthSix2023, gregoryAdvancingGlobalSea2026}. Generally, models may be improved by employing a multi-scale decomposition, such as from mrCOSTS, to diagnose the predictable and unpredictable aspects of the system and, by doing so, identify strategies for improving predictive models. 

The IceDMD framework has clear limitations. The bagging-derived uncertainty bands underestimate the true forecast uncertainty, particularly near seasonal extrema where non-stationary modes dominate (Fig. \ref{fig:dmd-forecast-eval}) and for individual locations (Fig. \ref{fig:pred-dmd-spatial+point}). The model also cannot reproduce sharp transitions between melting and freezing seasons. These two limitations are likely interrelated as a consequence of regularising the predictive model to capture the annual cycle and secular trend at the expense of higher-frequency and spatially-localised processes. Future work will address how to generate uncertainty quantification for practical applications. Most consequentially, IceDMD can only predict trends already present in the training window, i.e. it could not forecast the 2014--2017 event as this was driven by non-stationary interannual modes that are only partially resolved within the training window. It is for this reason that we targeted the period in which the secular trend emerged as a dominant source of variability. Against these limitations stand substantial practical advantages. The model trains in seconds on commodity hardware and is directly interpretable since each mode corresponds to the physical patterns uncovered here. Finally, even though the uncertainty was underestimated, the inclusion of any uncertainty estimate is an advantage since these were lacking in previous statistical S2I models \citep{massonnetSIPNSouthSix2023}. 

These strengths and limitations of IceDMD naturally suggest themselves towards future work incorporating DMD into more complex models.  In such a system DMD could form the ``backbone" of a predictive system by modelling the larger amplitude, stationary patterns of a system, thereby freeing up another method, such as a deep-learning, to focus on predicting the non-stationary patterns of the system. Such a system would naturally compliment the strengths of each approach while simultaneously mitigating their weaknesses. A predictive DMD model has stable long-term rollout and the capability of, when appropriately regularised, trivially capturing large-amplitude, stationary signals while deep-learning models typically struggle to produce stable long-term rollout but excel at learning finer scale patterns. Similar hybrid approaches in coupling physically-based and deep-learning models have already successfully improved sea ice forecasts \citep{gregoryAdvancingGlobalSea2026}. A hybrid DMD-deep learning approach could potentially enable even more powerful predictive models. Thus, we speculate a hybrid approach could use IceDMD's physically grounded, cheap baseline with multi-scale diagnostics to target what deep learning does best: modelling the non-stationary, high-frequency components that IceDMD deliberately excludes. Such a system may be able to overcome predictive barriers that typify S2I forecasting, for instance for Arctic sea ice \citep{bonanSpringBarrierRegional2019}, since the DMD backbone may be able to predict seasonal changes through the barrier. 

Antarctic sea ice entering a new climate regime is a natural test case for S2I forecasting. The combined analytical and predictive framework we present here, IceDMD, offers an interpretable, computationally negligible baseline against which more complex forecasting systems should be evaluated. This framework is likely generalizable beyond Antarctic sea ice and could provide progress in building seasonal forecasts for other parts of the earth system, as it relies on diagnosing the repeatable, and therefore predictable, patterns found within these complex systems.

\section{Methods}\label{methods}

\subsection{Sea Ice Data and Processing}

The analysis uses the publicly available Ocean and Sea Ice Satellite Application Facility (OSI SAF) Antarctic dataset, spanning from 1979 through 2024 and provides daily sea ice observations from passive-microwave sensors with atmospheric corrections applied \cite{lavergneVersion2EUMETSAT2019, EUMETSAT_SAF_SIC_v2}. Anomalies were computed by subtracting the average value for each individual month over the 40-year period (1984–2024) from the values for that same month in any specific year \citep{parkinson40yRecordReveals2019}. This method retains interannual variability but removes the annual cycle, effectively highlighting deviations from the long-term mean for each month. For fitting with DMD and mrCOSTS, we represent SIC as flattened images arranged into sequential snapshots, defined as ${\bf{x}}(t)$ (Fig. \ref{fig:overview}b). Sea Ice Area (SIA) was calculated as the spatial integral of SIC over a specified region of interest.

\subsection{Predictive DMD Model}
The predictive DMD model was fit to daily SIC snapshots, ${\bf{x}}(t)$, using the variable projection DMD technique \citep{ askhamVariableProjectionMethods2018} using Eq. \ref{eq:DMDsolution}. DMD is limited to fit only a specified number of modes, $r$. In practice, these modes come as conjugate pairs for which the imaginary component of the an individual mode's solution perfectly cancels out with its conjugate pair, resulting in a strictly real $\tilde{\mathbf{x}}$ \citep{lapoPhasorNotationDynamic2025}. Imposing strict conjugate pairs in the DMD solution results in odd rank modes that returns strictly real results for $\tilde{\mathbf{x}}$ with time dynamics that are either pure growth or decay and thus without oscillatory dynamics. We set the rank to $r=5$ and 2 years of historical data were used to fit DMD.

We used the BOP-DMD (Bagging, Optimized DMD) implementation available as part of PyDMD package \cite{askhamVariableProjectionMethods2018, sashidharBaggingOptimizedDynamic2022,ichinagaPyDMDPythonPackage2024} with stability constraints of no exponential growth (i.e., $\text{Re}(\omega)\leq0$) and strict conjugate pairs \citep{lapoPhasorNotationDynamic2025}. To quantify forecast uncertainty, we implement statistical bagging \cite{sashidharBaggingOptimizedDynamic2022}. By constructing multiple DMD models from randomly sampled subsets of ${\bf{x}}(t)$, we obtain an ensemble estimate of SIC dynamics. This bagging operation improves numerical stability and provides confidence intervals for our predictions. We used 100 bags with a sample size consisting of 20\% of the data. This sample size was chosen as larger sample sizes resulted in strongly underdispersive ensembles in the training split while smaller sample sizes resulted more dispersive ensembles that could not capture the secular trend (e.g., $Re(\lambda) = 0$ in Eq. \ref{eq:DMDsolution}). The bagging was performed external to PyDMD using PyDMD's BOPDMD optimisation as the core fitting routine for each member in order to retain individual ensemble members for uncertainty quantification.

\subsection{mrCOSTS}\label{sect:methods-mrcosts}

The mrCOSTS is a DMD variant capable of characterizing non-stationary, translating, multi-scale spatiotemporal features \cite{lapoMethodUnsupervisedLearning2025}. It operates by fitting data in sliding temporal windows of a specific time scale and separating the dynamics into low and high-frequency components. The low frequency component is passed to the next decomposition level with a larger window size that can resolve more of the low frequency dynamics. Upon completion, the dynamics are categorized again into frequency bands in order to capture dynamics leaked between decomposition levels, resulting in coherent spatial modes described by relatively narrow bands of time dynamics. 

The resulting bands are denoted using the index $p$, which is comprised by a subset of the ranks ($j$), time windows ($k$), and decomposition levels ($\ell$) of DMD modes that exist within the frequency band, defined as $(j, k, \ell) \in p$. The reconstruction of a band in this situation can be notated as
\begin{equation}\label{eq:summed-amplitude}
\tilde{\mathbf{x}}_p(t)=\sum_{(j, k, \ell) \in p} b_{j, k, \ell} e^{\omega_{j, k, \ell}t} \boldsymbol{\phi}_{j, k, \ell}.
\end{equation}
The global reconstruction of the original data is found by summing over the individual frequency bands, 
\begin{equation}
\tilde{\mathbf{x}}(t)\approx\sum_{p=0}^K \tilde{\mathbf{x}}_p(t) +  \mathbf{c}_{0, p}, \label{eq:global-reconstruction}
\end{equation}
where $\mathbf{c}_{0, p}$ are the mean values of the individual windows from the final decomposition level and $K$ is the number of total bands found. The low-frequency band from the last decomposition level may not contain well-resolved dynamics since it describes time scales longer than the largest decomposition level's window width. This band is referred to as the background mode, or $\bf{\tilde{x}}_{bgd}$. Reconstructions of a specific frequency bands can be found using Eq. \ref{eq:global-reconstruction} and summing over the desired $p$ and omitting the background value.

\begin{figure}[h]
    \centering
    \includegraphics[width=1\linewidth]{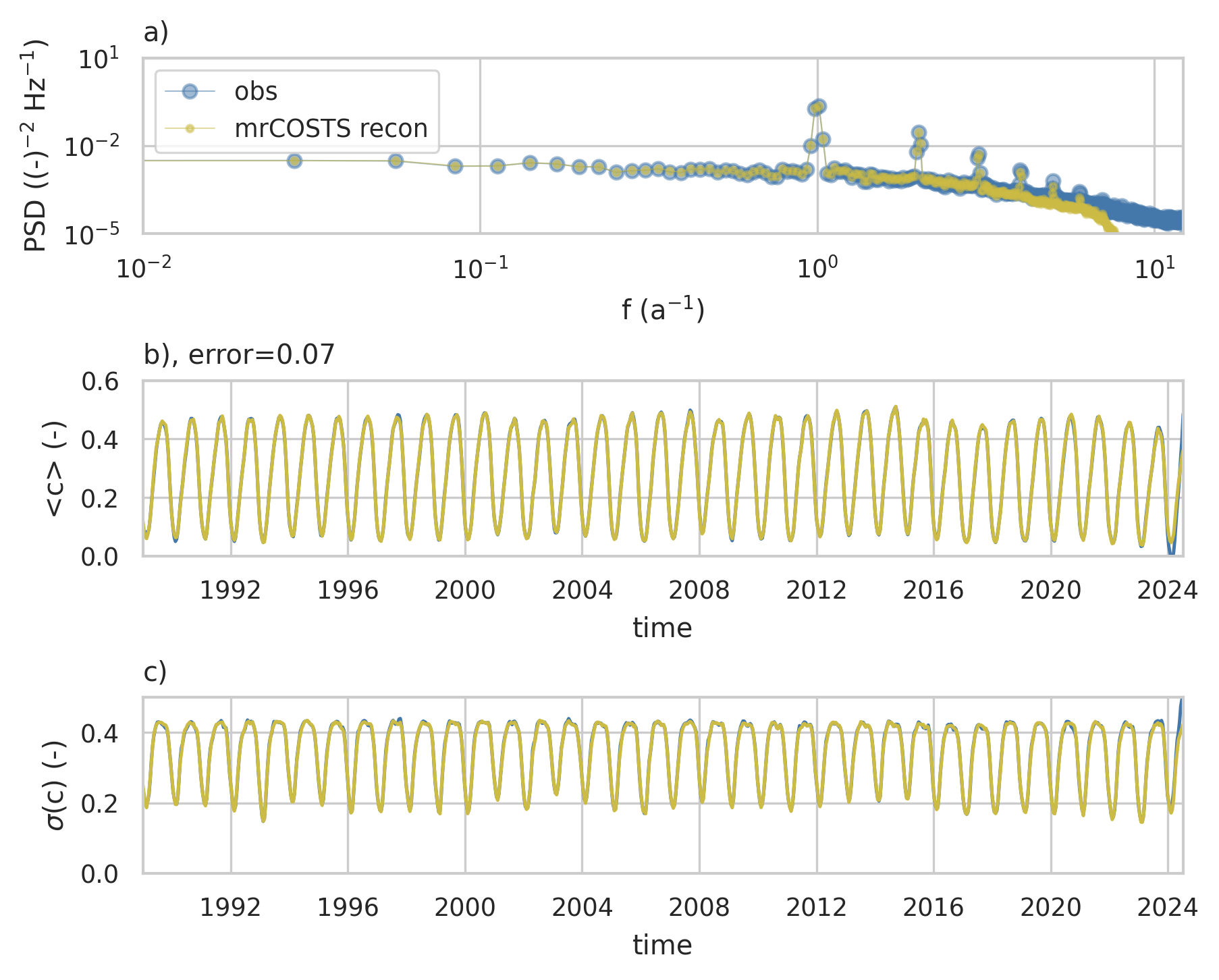}
    \caption{The mrCOSTS diagnostic robustly reconstructs the SIC data. (a) The power spectral density of the observed (blue) and mrCOSTS reconstruction (yellow) SIC. The low order statistical moments of (b) mean SIC and (c) standard deviation of SIC are well reconstructed.}
    \label{fig:mrcosts-eval}
\end{figure}

The mrCOSTS diagnostic is evaluated by reconstructing the original input data. We can accurately reconstruct SIC with an error of 0.07, with most of the error caused by not capturing the highest frequencies (Fig. \ref{fig:mrcosts-eval}a) and the known biases at the edges of the time domain, which are analogous to the cone-of-influence from continuous wavelet transforms (Fig. \ref{fig:mrcosts-eval}b,c). All components of the SIC with a period of 2 months or longer are well captured, indicating a successful decomposition of the SIC dynamics. Hyperparameter tuning was performed in order to better capture the shortest time scales, but the qualitative results from mrCOSTS did not significantly change between hyperparameter sets. The hyperparameters employed in this study can be found as part of the publicly available code accompanying this study (see Code Availability).

\section*{Acknowledgements}
We thank Ed Blockley (Met Office Hadley Centre) and Caroline Holmes (British Antarctic Survey) for their helpful comments when writing the manuscript. P.Y. was supported by the Ecosystem Leadership Award under the EPSRC Grant EP/X03870X/1 and The Alan Turing Institute, particularly the Turing Research Fellowship scheme under that grant. J.N.K. acknowledges support from the National Science Foundation AI Institute in Dynamic Systems grant 2112085. K.L. was funded by the Austrian Science Fund (FWF) [10.55776/ESP214]. O.S., L.v.Z and J.S.H. was funded by The Alan Turing Institute.

\section*{Data Availability}
The OSI SAF sea ice concentration data (v2) used in this study are publicly available from the EUMETSAT user portal (\url{http://dx.doi.org/10.15770/EUM_SAF_OSI_0008}), and the post-processed temporally complete daily data is available on Zenodo (\url{https://doi.org/10.5281/zenodo.19679650}).

\section*{Code Availability}
The mrCOSTS and optimised DMD implementations are available as part of the PyDMD Python package (\url{https://github.com/PyDMD/PyDMD}). Code for reproducing the work in this manuscript is available through \url{https://github.com/alan-turing-institute/icedmd-antarctic-paper}.

\section*{Author Contributions}
P.Y. and K.L. contributed equally to this work. P.Y. co-conceived the study, led development of the predictive methodology, contributed to the analysis, and contributed to writing the manuscript. K.L. performed the mrCOSTS analysis, predictive DMD uncertainty analysis, led writing of the manuscript, and contributed to methodology, analysis, and interpretation. J.S. co-conceived the study and contributed to the development of the predictive methodology and analysis. O.S. contributed to computational implementation. L.V.Z. contributed to data processing and analysis. J.S.H. provided domain expertise on Antarctic sea ice and contributed to interpretation. J.N.K. supervised the DMD methodology development and contributed to manuscript preparation.

\section*{Competing Interests}
The authors declare no competing interests.

\bibliographystyle{unsrtnat}
\bibliography{bibliography1.bib}

\end{document}